\documentclass[journal]{IEEEtran}
\usepackage{amsmath,amssymb,amsfonts}
\usepackage{algorithm}
\usepackage{algorithmic}
\usepackage{array}
\usepackage[caption=false,font=normalsize,labelfont=sf,textfont=sf]{subfig}
\usepackage{textcomp}
\usepackage{stfloats}
\usepackage{subfig}
\usepackage{url}
\usepackage{bm}
\usepackage{verbatim}
\usepackage{graphicx}
\usepackage{cite}
\usepackage{booktabs}
\usepackage{xcolor}
\begin{document}

\newcounter{proposition}
\setcounter{proposition}{0}
\newenvironment{proposition}[1][]{\refstepcounter{proposition}\textbf{Proposition \arabic{proposition}:} \rmfamily}{\medskip}

\title{Hybrid Near-Field and Far-Field Localization with Multiple Reconfigurable Intelligent Surfaces}

\author{Weiqiao Zhu,
        Mengyuan~Cao,~\IEEEmembership{Student Member, IEEE},
        Yang Yang,
        Haobo~Zhang,~\IEEEmembership{Member, IEEE},
        Weijun Hao,
        Xiaofei Jia,
        and Hongliang~Zhang,~\IEEEmembership{Member, IEEE}.
		\thanks{
            W. Zhu, Y. Yang, W. Hao, and X. Jia are with Institute of Computing Technologies, China Academy of Railway Sciences Corporation Limited, Beijing 100081, China (e-mail: \{zhuweiqiao, yangyang, haoweijun\}@rails.cn, carsrails@gmail.com).
            
			M. Cao and Hongliang Zhang are with School of Electronics, Peking University, Beijing 100871, China (e-mail: \{caomengyuan, hongliang.zhang\}@pku.edu.cn). 

            Haobo Zhang is with the School of Electronic and Computer Engineering, Peking University Shenzhen Graduate School, Shenzhen 518055, China (e-mail: haobo.zhang@pku.edu.cn).
			}
	}



\maketitle

\begin{abstract}
Localization using multiple base stations~(BSs) has gained much attention for its advantage in localization accuracy. However, the performance of the multi-BS system suffers from its limited number of antennas. To solve the above issue, we propose to use reconfigurable intelligent surfaces~(RIS) serving as antennas. Existing localization methods enabled by multiple RISs mainly focus on the far-field~(FF) region of each RIS. As the scale of RIS increases, the near-field~(NF) region of each RIS expands, where FF methods struggle to achieve high localization accuracy. In this letter, a hybrid NF and FF localization method aided by multiple RISs is proposed. In such scenarios, achieving user localization and RIS optimization becomes challenging due to the high complexity caused by the exhaustive search through all candidate locations to match the signals. Moreover, the interference from multiple RISs degrades the localization accuracy. To address this challenge, we propose a two-phase localization method that first estimates the relative locations of the user to each RIS and fuses the results to obtain the estimation. This approach reduces the complexity by decreasing the number of candidate locations considered in each step. Also, we introduce a constraint in the RIS optimization problem that limits the sidelobe levels directed towards other RISs, effectively minimizing inter-RIS interference. The effectiveness of the proposed method is verified through simulations.
\end{abstract}


\section{Introduction}
The growing number of location-dependent services, such as navigation systems, user monitoring, and autonomous vehicles, has increased the need for precise localization. This surge in demand has inspired the exploration of innovative localization methods. Benefiting from the deployment of multiple base stations~(BSs), the diversity and coverage can be enhanced, thus improving the localization accuracy\cite{b75}. However, multi-BS systems suffer from the limited number of antennas due to the power constraint and hardware cost\cite{b76}, which restricts the performance. To solve this problem, we employ the reconfigurable intelligent surface (RIS) to serve as antennas of the BS. Specifically, RIS is composed of multiple sub-wavelength nearly-passive scattering elements, and can manipulate electromagnetic (EM) fields and generate desirable beams with high precision due to its massive number of tunable elements~\cite{b72}.

In the literature, some existing works have considered the localization systems aided by multiple RISs. In \cite{b69}, the authors considered a multi-reflection wireless environment, and a two-stage user localization method based on the atomic norm minimization and least-squares line intersection is proposed. The authors in \cite{b70} proposed a method based on the time of arrival, angle of arrival, and the time difference of arrival to locate the user in a non-line-of-sight environment. In~\cite{b78}, the authors proposed a recursive localization scheme with an iterative RIS selection strategy. However, these methods primarily rely on the traditional plane wave approximation. While this approximation is valid for the far-field (FF) region, the large aperture of RISs significantly extends their near-field (NF) region. In the NF region, the plane wave approximation becomes inadequate for precise signal propagation modeling, leading to compromised localization accuracy for existing methods. To address this issue, the spherical wave model should be adopted for the NF. Practically, users may be located in the NF or the FF region of the RIS. Therefore, a hybrid localization scheme is necessary to accommodate both regions.

In this letter, a localization scheme is proposed where a BS and multiple RISs are coordinated to serve multiple users in the hybrid NF and FF regions. Specifically, multiple single-antenna users transmit narrowband signals. The BS processes the received signal reflected by the selected RISs of each user to estimate user locations. Based on these estimates, the RIS selection and phase shifts are optimized to improve the localization performance. The system then iteratively alternates between localization and RIS optimization.

Several challenges have arisen for the proposed localization scheme. \textit{First,} for multiple RISs, the reflection signals from each RIS at different locations require an exhaustive search across all candidate locations to match the received signals, resulting in high computational complexity. In response to this problem, we design a two-phase localization algorithm that first estimates the relative locations of the user to each RIS and fuses the results to obtain the estimation. This approach reduces the complexity by decreasing the number of candidate locations considered in each step. \textit{Second,} the received signal could contain the signal reflected multiple times by different RISs, which causes interference between RISs. To tackle this problem, we add a constraint in the RIS optimization problem to limit the sidelobe levels directed towards other RISs, and design an algorithm based on the alternating direction method of multipliers to solve the formulated problem. 

\begin{figure}
	\centering
	\includegraphics[scale=0.33]{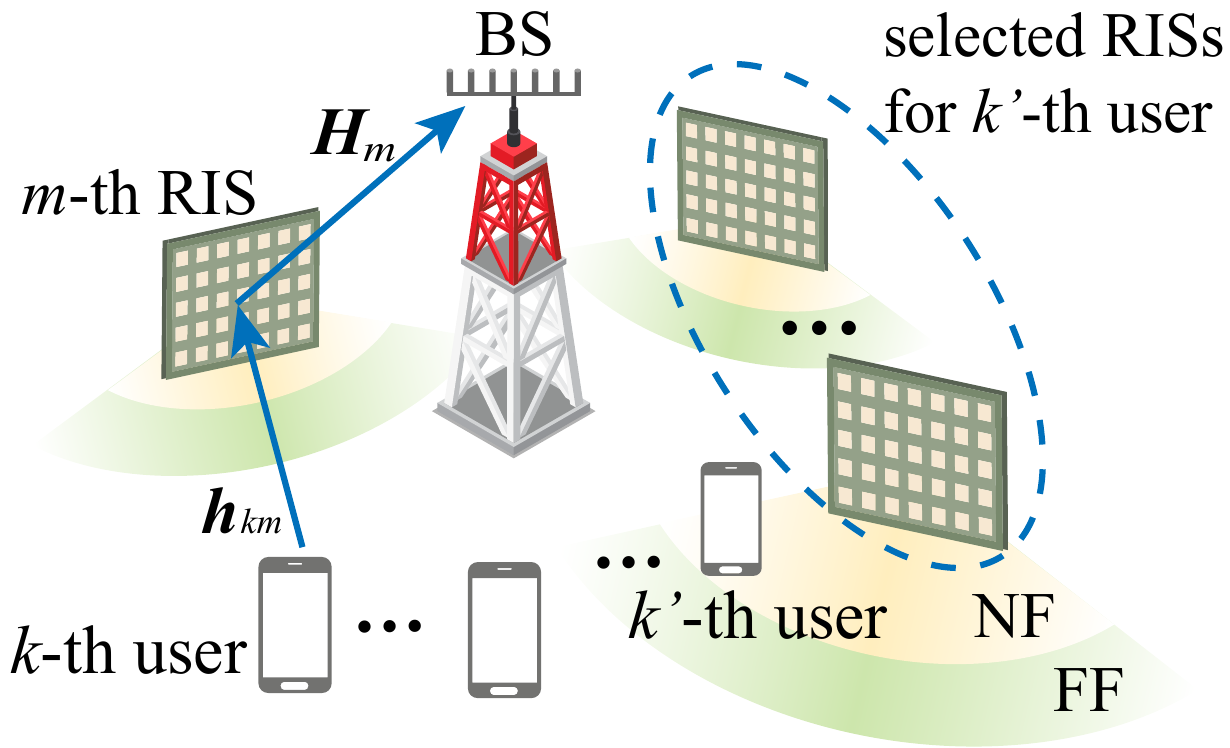}
	\caption{Hybrid NF and FF localization aided by multiple RISs.}
	\label{fig:sce}
\end{figure}

\section{System Model}
\label{section:sm}

In this section, the localization scenario is first described, followed by an introduction to the signal model and the proposed localization protocol.

\subsection{Localization Scenario}

The system architecture, depicted in Fig. \ref{fig:sce}, comprises a BS equipped with $V$ antennas, $K$ single-antenna users, and $M$ RISs. Each RIS contains $N = N_1 \times N_2$ programmable reflecting elements with tunable phase shifts.

During the localization, the RISs that offer more reliable connections are selected to locate each user. Each user transmits a narrowband scalar symbol $s_k$, which is then reflected by all RISs. The BS separates the signal reflected by each RIS by digital beamforming\cite{b69}, and performs localization for each user using the signals reflected by its selected RIS. To prevent interference among multiple users, frequency division multiplexing~(FDM) is employed for different users. Note that each user can be in either the NF or the FF region of each RIS, and which region a user locates in is unknown.

\subsection{Channel Model}

Based on the cascaded channel model, the channel between the $k$-th user and the BS via the $m$-th RIS can be given by~\cite{b80}
\begin{align}
	\bm{h}_{km}(\bm{p}_k^m, \bm{Q}^m) = \bm{H}_{m}^T \text{diag}\{\bm{\beta}_{m}\} \bm{h}_{km}^t(\bm{p}_k^m, \bm{Q}^m),
\end{align}
where $\bm{p}_{k}^{m} = [r_k^m, \theta_k^m, \phi_k^m]^T$ represents the relative location of the $k$-th user with respect to the $m$-th RIS, with $r_k^m, \theta_k^m, \phi_k^m$ being the range, azimuth, and elevation angle, respectively. Let $\bm{Q}^m = [\bm{q}_1^m, ... \bm{q}_L^m] \in \mathbb{C}^{3\times L}$ denote the locations of scatters, where $L$ is the total number of scatters and $\bm{q}_l^m$ is the location of the $l$-th scatter. $\bm{\beta}_{m}=[\beta_{1}^{m},...,\beta_{N}^{m}]^T \in \mathbb{C}^{N \times 1}$ represents the phase shift of the $m$-th RIS, where each element $\beta_{n}^{m}$ corresponds to the phase shift of the $n$-th RIS element. $\bm{H}_{m} \in \mathbb{C}^{N\times V}$ represents the channel from the $m$-th RIS to the BS, and $\bm{h}_{km}^t(\bm{p}_k^m, \bm{Q}^m) \in \mathbb{C}^{N\times 1}$ represents the channel between the $m$-th RIS and the $k$-th user, as visualized in Fig.~\ref{fig:sce}. Here, we assume the RIS-BS channel $\bm{H}_{m}$ and RIS phase shift $\bm{\beta}_{m}$ are perfectly known.

The user-RIS channel can be modeled by\cite{b36}
\begin{align}
    \bm{h}_{km}^t(\bm{p}_k^m, \bm{Q}^m) = \bm{h}_{km0}(\bm{p}_k^m) + \sum_{l=1}^L \bm{h}_{kml}(\bm{p}_k^m, \bm{q}_l^m),
\end{align}
where $\bm{h}_{km0}$ is the direct path between the user and RIS. $\bm{h}_{kml}$ is the $l$-th scattering path between the $k$-th user and the $m$-th RIS. The RIS-user direct path expressions for both NF and FF users are presented separately in the following.

\textbf{Channel Models for the NF Region}:
When the user is located in the NF region of the $m$-th RIS, we employ the spherical wave model to characterize the signal received by the RIS. Thus, the direct path between the $k$-th user and the $m$-th RIS can be modeled as\cite{b6}
\begin{align}
	\label{nfdirect}
	\bm{h}_{km0}(\bm{p}_k^m) = \alpha_{km} \bm{b}(\bm{p}_k^m), \ \bm{p}_k^m \in \bm{D}_{NF}^m,
\end{align}
where $\alpha_{km}$ denotes the channel gain coefficient~\cite{b20}. $\bm{b} (\bm{p}_k^m) \! = \! \left[\exp\!\left(-j\frac{2\pi}{\lambda}d_{km1}^t\right),..., \exp\!\left(-j\frac{2\pi}{\lambda}d_{kmN}^t\right)\right]^T$ is the NF steering vector,
where $d_{kmn}^t = \Vert \bm{p}_k^m - \bm{p}_n^m \Vert$, with $\bm{p}_n^m = (0, y_n, z_n)$ represents the local coordinates of the $n$-th RIS element. $\bm{D}_{NF}^m$ represents the NF region of the $m$-th RIS\cite{b77}. 

\textbf{Channel Models for the FF Region}:
\label{subsubsection:ff}
For FF users, the plane wave model is employed to characterize the received signal, and the direct user-RIS path is modeled by~\cite{b34}
\begin{align}
	\label{ffdirect}
	\bm{h}_{km0}(\bm{p}_k^m) = \alpha_{km} \bm{a}(\theta_k^m, \phi_k^m), \ \bm{p}_k^m \in \bm{D}_{FF}^m,
\end{align}
where $\bm{D}_{FF}^m$ represents the FF region of the $m$-th RIS. $\bm{a}(\theta, \phi)$ is the steering vector in the FF region given in\cite{b77}.

\subsection{Localization Protocol}
\label{subsection:LP}
\begin{figure}
	\centering
	\includegraphics[scale=0.25]{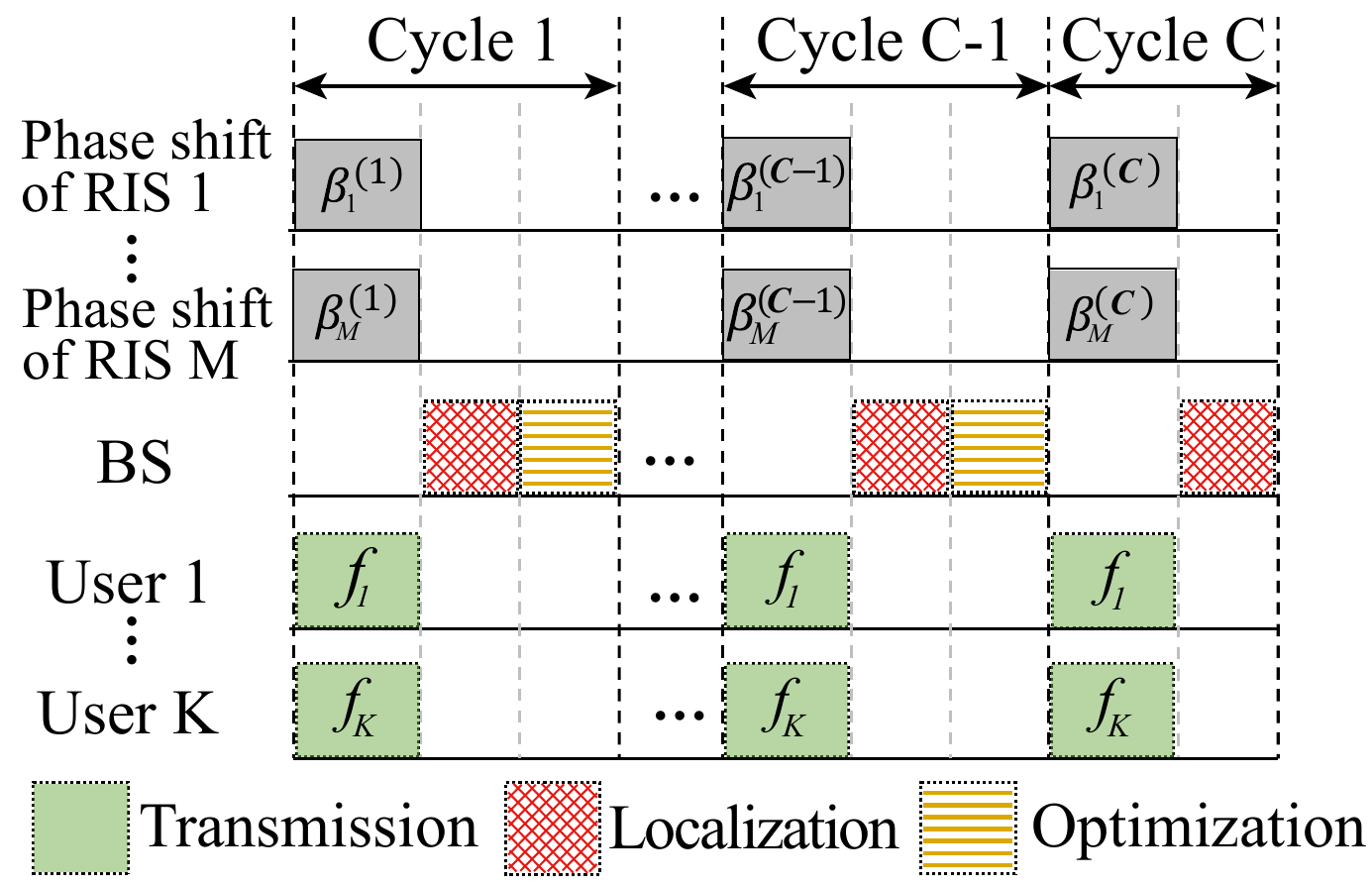}
	\caption{Hybrid NF and FF user localization protocol enabled by multiple RISs.}
	\label{fig:2}
\end{figure}

The localization process consists of $C$ iterative cycles, where each cycle executes three steps: transmission, localization, and optimization. Fig. \ref{fig:2} provides a schematic overview of the localization protocol.

\subsubsection{Transmission}
In this step, the users broadcast signals to the surroundings, which are then reflected by the RISs and received by the BS. Let $g^{(c)}_{mk}$ denote the signal transmitted by $k$-th user and reflected by the $m$-th RIS in the $c$-th cycle.

\subsubsection{Localization}
\label{subsubsection:localization}
In the next step, the BS estimates each user's location using the received signals reflected by the selected RISs, denoted by $\hat{\bm{P}}^{(c)} =[\hat{\bm{p}}^{(c)}_1, \cdots, \hat{\bm{p}}^{(c)}_K]$. The localization algorithm is introduced in Sec.~\ref{section:location}.

\subsubsection{Optimization}

In this step, the BS first optimize the RIS selection for each user, and then determines the optimal RIS phase shifts of the $(c+1)$-th cycle $\bm{\beta}^{(c+1)}_m$ based on $\hat{\bm{P}}^{(c)}$ and the RIS selection results. Note that this step is not executed in the last cycle. The proposed RIS optimization is detailed in Sec.~\ref{section:prob}.

\section{Multiple RISs-enabled Localization}
\label{section:location}
In this section, we formulate the localization problem aided by multiple RISs. A localization algorithm is then designed to effectively solve the formulated problem.

\subsection{Overall Problem}
We formulate the localization problem by minimizing the $l_2$-norm residuals between received signals and those reconstructed from location estimates. The localization problem in the $c$-th cycle is given as\cite{b20}
\begin{subequations}
	\begin{flalign}
		\!\!\text{P1}\!:\!  \mathop{\min}_{\hat{\bm{P}}^{(c)}}  \sum_{m \in \mathcal{M}_k}&\left\Vert \bm{g}_{mk}^{(c)} -\bm{w}_m\bm{h}_{km}(\bm{p}_k^m, \bm{Q}^m)s_k \right\Vert_2^2 \!, \!\! \\
		\!\!&s.t. \ \ \tilde{\bm{p}}_k^m \in \bm{D},
	\end{flalign}
\end{subequations}
where $\mathcal{M}_k$ is the set of the selected RISs for the $k$-th user. $\bm{g}_{mk}^{(c)} = [g^{(1)}_{mk}, ..., g^{(c)}_{mk}]$ is the received signal for the $k$-th user for $c$ cycles. $\bm{w}_m$ is the digital beamforming vector directed towards the $m$-th RIS. $\bm{D}$ is the whole search area. Note that the user-RIS paths through multiple RIS reflections will be minimized in the RIS phase shift optimization. Therefore, the multi-reflection paths are ignored in the problem formulation.

However, if the received signals reflected by all RISs are used simultaneously to locate the user, it is necessary to sample the entire space and compare the received signals with the signals at each sampling point one by one, which leads to high complexity. To address this issue, a two-phase localization method is proposed. Specifically, we first independently locate the user using the reflected signals from a single RIS, and then fuse the localization results from all RISs.  

\subsection{Local Localization}
\label{subsec:dis}

First, we decouple (P1), and local localization problem for the $k$-th user via the reflection of the $m$-th RIS is given as
\begin{subequations}
	\begin{flalign}
		\label{6a}\!\!\text{P1'}\!:\!  \mathop{\min}_{\tilde{\bm{p}}_{k}^m}  &\left\Vert \bm{g}_{mk}^{(c)} -(\bm{B}_m^{(c)})^T \text{diag} \{\bm{w}_m\bm{H}_m^T \}  \bm{h}_{km}^t s_k \right\Vert_2^2 \!, \!\! \\
		\!\!&s.t. \ \ \tilde{\bm{p}}_k^m \in \bm{D},
	\end{flalign}
\end{subequations}
where $\bm{B}_m^{(c)} = [\bm{\beta}_m^{(1)},...,\bm{\beta}_m^{(c)}]$ is the phase shift of the $m$-th RIS for $c$ cycles. To solve (P1'), we propose an algorithm based on orthogonal matching pursuit~(OMP), a classic sparse recovery algorithm. Following the method in \cite{b77}, we sample the NF and FF regions of each RIS. To account for the different signal models in the NF and FF regions, we sample the angle and range for the NF region, and only the angle for FF region. The sampling spacings for range, azimuth, and elevation angles are $\Delta R, \Delta \theta, \Delta \phi$, respectively. The sampled locations are given by $\bm{Z} = [\bm{Z}_{near}, \bm{Z}_{far}]$.
Then, the steering vectors of the sampled locations form the atom channels $\bm{F} = [\bm{F}_{near},\bm{F}_{far}]$, where
\begin{align}
    \bm{F}_{near} &= [\bm{b}(r_1^m, \theta_1^m, \phi_1^m),...,\bm{b}(r_{S_1}^m, \theta_{S_1}^m, \phi_{S_1}^m)], \\
    \bm{F}_{far} &= [\bm{a}(\theta_1^m, \phi_1^m),...,\bm{a}(\theta_{S_2}^m, \phi_{S_2}^m)].
\end{align}
where $[r_i^m, \theta_{i}^m, \phi_i^m]$ and $[\theta_{i}^m, \phi_{i}^m]$ are the $i$-th sampled NF or FF location. $S_1$ and $S_2$ are the total number of sampled NF and FF locations. 
Then, (P1') can be approximated as\cite{b77}
\begin{subequations}
	\begin{flalign}
    \label{abc}
		\!\!\!\!\text{P1''}:  \mathop{\min}_{\bm{u}_k}  &\left\Vert \bm{g}_{mk}^{(c)} - (\bm{B}_m^{(c)})^T \text{diag} \{\bm{w}_m\bm{H}_m^T \}\bm{F}\bm{u}_k s_k \right\Vert_2^2\!,\!\!\!\\
		\label{bcd}
        \!\!\!\! s.t. &\ \ \Vert \bm{u}_k \Vert_0 = L+ 1,
	\end{flalign}
\end{subequations}
where $\bm{u}_k$ is the amplitude of the atom channels $\bm{F}$ for the $k$-th user. Hence, (P1') can be solved by OMP. Specifically, suppose $\tilde{\bm{u}}_{k}$ is the solution of (P1'), and the index of its biggest element is $i$, the estimated location of the $k$-th user can be given by
\begin{align}
    \tilde{\bm{p}}_{k}^m=
        \begin{cases}
           [r_i^m, \theta_{i}^m, \phi_i^m]^T,  &(i \leq S_1), \\
           [\theta_{i-S_1}^m, \phi_{i-S_1}^m]^T, &(i > S_1),
        \end{cases}
\end{align}

\subsection{Location Fusion}
\label{subsec:res}

After the BS independently estimate the local locations of all the users, the estimation results are fused to obtain the global estimation. We define the fusion loss as the sum of squared relative distances between the global estimated location and all the local estimated locations. This choice of loss function is motivated by penalizing large deviations between the global and local estimates. Note that the local estimation results in NF and FF regions have different forms, i.e., with or without range information. To accommodate both the NF and FF estimation results, we need to separately calculate the relative distances for each case.

\textbf{NF case}:
The relative distance for NF users is given by $d_{mk} = \Vert \hat{\bm{p}}_k - \tilde{\bm{p}}^{m}_{k}\Vert$,
where $\hat{\bm{p}}_k$ is the global estimated location of the $k$-th user, and $\tilde{\bm{p}}_{k}^{m} = [\tilde{r}_{mk}, \tilde{\theta}_{mk}, \tilde{\phi}_{mk}]^T$ is the location estimated using the received signal reflected by the $m$-th RIS.

\textbf{FF case}:
The FF localization result only contains angle information and corresponds to a ray in 3D space. Thus, we define the FF relative distance as the minimum distance from the estimated global location to the ray, which is given as
\begin{align}
	d_{mk} = \Vert \overrightarrow{A_m P_k} \times  \overrightarrow{A_m B_{mk}}\Vert,
\end{align}
where $\overrightarrow{A_m P_k}$ is the vector from the center of the $m$-th RIS to the estimated global location. $\overrightarrow{A_m B_{mk}}$ is the unit vector parallel to the estimated FF direction $\tilde{\bm{p}}_k^m = [\theta_k^m, \phi_k^m]$.

For simplicity, we consider a rectangular space, whose boundaries along the three coordinate axes are given by $x_{\pm}$, $y_{\pm}$, and $z_{\pm}$. Therefore, the fusion problem can be formulated as\cite{b79}
\begin{subequations}
    \begin{align}
    	&\text{P2:} \min_{\hat{\bm{p}}_k} \sum_{m \in \mathcal{M}_k} d_{mk}^2(\hat{\bm{p}}_k)\\
          s.t.&\ \ \hat{\bm{p}}_k \in [x_{-},x_+] \times [y_{-},y_+] \times [z_{-},z_+].
    \end{align}
\end{subequations}
Then, the solution can be derived by quadratic programming.

\section{RIS Optimization}
\label{section:prob}

This section focuses on optimizing both RIS selection and phase shift configuration. We begin by introducing the RIS selection strategy, then formulate the phase shift optimization problem and derive a solution based on the alternating direction method of multipliers (ADMM).

\subsection{RIS Selection Optimization}
The accuracy and reliability of the localization results can be improved by choosing the most effective RISs to serve each user. Since the channel gain between the RIS and the user is inversely proportional to their relative distance $\hat{r}_{km} = \Vert \bm{q}_m - \hat{\bm{p}}_k\Vert$, where $\bm{q}_m$ is the location of the $m$-th RIS, we can use the relative distance as the indicator. Therefore, after obtaining the global estimated locations for all users, we select the $L$ RISs with the minimum distances for each user~\cite{b66}.

\subsection{Phase Shift Optimization}

\subsubsection{Problem Formulation}

In the optimization step, the BS optimizes the RIS phase shifts based on the global localization results. The primary metric is the Cramér-Rao Bound~(CRB), providing a theoretical lower bound for estimation error. Additionally, to minimize the interference between RISs, the RIS sidelobe amplitude directed towards other RISs is minimized. Hence, we aim to simultaneously minimize RIS sidelobe levels and the sum of the estimation CRBs for all users. Thus, the optimization problem is formulated as follows\cite{b6}\cite{b59}
\begin{subequations}
	\begin{align}
		\text{P3}: &\mathop{\min}_{\bm{\beta}_{m}^{(c+1)},\eta}  \epsilon \log \eta +  \sum_{k\in \mathcal{K}_m} tr \left(\bm{J}_{mk}^{-1} \right), \\ 
		 s.t. \ &|\beta_{mn}^{(c+1)}|=1,\ \forall n,  \label{eq:sub27b}\\
		 & \vert \bm{w}_m\bm{H}_{m}^T\text{diag}\{\bm{\beta}_{m}^{(c+1)}\}\bm{a}(\theta_{im},\phi_{im}) \vert_2^2 \leq \eta, i \neq m,\label{eq:sub27c}
	\end{align}
\end{subequations}
where $\mathcal{K}_m$ represents the set of users that selected $m$-th RIS. $\bm{J}_{mk}$ is the Fisher information matrix of the user location $\bm{p}_k$, which is a function of the RIS phase shifts $\bm{\beta}_{m}^{(c+1)}$ and can be calculated according to\cite{b77}. $\epsilon$ is the penalty factor to balance the impact of the maximum allowed sidelobe amplitude $\eta$. (\ref{eq:sub27b}) is the constant-modulus constraint of the RIS phase shift. (\ref{eq:sub27c}) is the sidelobe suppression constraint, ensuring that the radiation pattern $\bm{w}_m\bm{H}_{m}^T\text{diag}\{\bm{\beta}_{m}^{(c+1)}\}\bm{a}(\theta_{im},\phi_{im})$ of the angle $(\theta_{im},\phi_{im})$ is below $\eta$. Here, each RIS is assumed to lie in the FF region of other RISs, and $\{(\theta_{im},\phi_{im})\}_{i\neq m}$ are the relative angles of other RISs for the $m$-th RIS. In this way, the path formed by multiple RIS reflections can be neglected.

\subsubsection{Algorithm Design}

Conventional RIS phase shift optimization algorithms, such as the complex circle manifold method, are designed to handle only the constant-modulus constraint (\ref{eq:sub27b}) and thus are unable to effectively solve (P3) due to the sidelobe constraint (\ref{eq:sub27c}). To address this issue, we propose an ADMM-based algorithm that iteratively handles each constraint to tackle the formulated problem. We introduce the auxiliary variable $\bm{f}=\{f_{i}\}_{i\neq m}$, which is defined as $f_{i} = \bm{w}_m\bm{H}_{m}^T\text{diag}\{\bm{\beta}_{m}^{(c+1)}\}\bm{a}(\theta_{im},\phi_{im})$. Then, the partial augmented Lagrangian of (P3) can be given by~\cite{b59}
\begin{align}
    &L_{\gamma} =  \epsilon \log \eta + \sum_{k \in \mathcal{K}_m}  tr \left(\bm{J}_{k}^{-1} \right) 
    + \dfrac{\gamma}{2} \sum_{i\neq m} (\vert f_{i} - \nonumber\\&\bm{w}_m\bm{H}_{m}^T\text{diag}\{\bm{\beta}_{m}^{(c+1)}\}a(\theta_{im},\phi_{im}) + \lambda_{i} \vert^2 -\vert \lambda_{i}\vert^2),
\end{align}
where $\gamma$ is penalty factors, $\lambda_{i}$ is scaled dual variable with $i\neq m$. Then, we iteratively solve the variables $\bm{\beta}_{m}^{(c+1)}$, $\bm{f}$, $\eta$ and $\{\lambda_{i}\}$. For simplicity, $\bm{\beta}_{m}^{(c+1)}$ is denoted as $\bm{\beta}$ in the following. The subproblems are given by
\begin{subequations}
    \begin{align}
        \!\!\bm{\beta}^{t+1} = &\arg \min_{\bm{\beta}} L_{\gamma}(\bm{\beta},\bm{f}^t,\eta^t,\lambda_{i}^t), \nonumber\\
        \label{20a}s.t.& \vert \beta_n \vert =1, \\
        \!\!\{\bm{f}^{t+1},&\eta^{t+1}\} = \arg \min_{\bm{f},\eta} L_{\gamma}(\bm{\beta}^{t+1},\bm{f},\eta,\lambda_{i}^t), \nonumber \\
        \label{20xb}s.t. &\vert f_{i} \vert^2 \leq \eta, \\
        \label{20c}\!\!\lambda_{i}^{t+1} = &\lambda_{i}^{t} +  f_{i}^{t+1} -\bm{w}_m\bm{H}_{m}^T \text{diag}\{\bm{\beta}^{t+1}\}\bm{a}(\theta_{im},\phi_{im}).
    \end{align}
\end{subequations}

\textbf{Solution to (\ref{20a})}:
Define $v_{i} = f_{i} + \lambda_{i}$ and $\bm{x}_{i}^T = \bm{w}_m\bm{H}_{m}^T\text{diag}\{\bm{a}(\theta_{im},\phi_{im})\}$, we can rewrite (\ref{20a}) as
\begin{subequations}
    \begin{align}
        \min_{\bm{\beta}} &\sum_{k \in \mathcal{K}_m}  tr \left(\bm{J}_{k}^{-1} \right) + \Vert \bm{v} - \bm{X}\bm{\beta}\Vert^2, \label{19a} \\
        &s.t. \vert \beta_n\vert =1, \ \forall n \label{xx1}
    \end{align}
\end{subequations}
where $\bm{v} = [v_{i}]$, and $\bm{X} = [\bm{x}_{i}], i \neq m$. Due to the constant modulus constraint (\ref{xx1}), we can employ the complex circle manifold~(CCM) algorithm to solve the above problem~\cite{b14}.

The main idea of CCM is to perform gradient descent in the complex circle manifold space. First, we compute the Euclidean gradient of the objective function and project it onto the manifold's tangent space to obtain the Riemannian gradient. The Riemannian gradient of the first and second terms in (\ref{19a}) are given in~\cite{b77} and~\cite{b59}, respectively. Then, the phase shift is updated along the Riemannian gradient, followed by retraction to guarantee that it remains on the manifold. This process is iteratively conducted until convergence.

\textbf{Solution to (\ref{20xb})}:
Define $\hat{f}_{i} = \bm{w}_m\bm{H}_{m}^T\text{diag}\{\bm{\beta}^{t+1}\}a(\theta_{im},\phi_{im}) - \lambda_{i}^t$ and ignore the irrelevant terms, we can translate the subproblem (\ref{20xb}) into
\begin{subequations}
    \begin{align}
        \min_{\bm{f},\eta} \ &\epsilon \log \eta + \dfrac{\gamma}{2}  \sum_{i\neq m} \vert f_{i} - \hat{f}_{i} \vert^2, \\
        &s.t. \vert f_{i} \vert^2 \leq \eta.
    \end{align}
\end{subequations}

The solution of $f_{i}$ can be given as\cite{b59}
\begin{align} 
    f_{i}=\begin{cases}\sqrt{\eta}\frac{\hat{f}_{i}}{\vert \hat{f}_{i}\vert}, & \text{if}\ \vert \hat{f}_{i}\vert > \sqrt{\eta},\\ 
    \hat{f}_{i}, & \text{otherwise}.\end{cases}
\end{align}

Then, the solution of $\eta$ can be given by
\begin{align}
\min_{\eta} \ \epsilon\log\eta+\frac{\gamma}{2}\sum_{i\neq m}\rho_{i}(\sqrt{\eta}-\vert \hat{f}_{i}\vert)^{2},
\end{align}
where $\rho_{i} = 0$ if $\vert \hat{f}_{i}\vert \leq \sqrt{\eta}$, otherwise, $\rho_{i} = 1$.

\begin{figure*}[t!]
	\centering
	\includegraphics[scale=0.42]{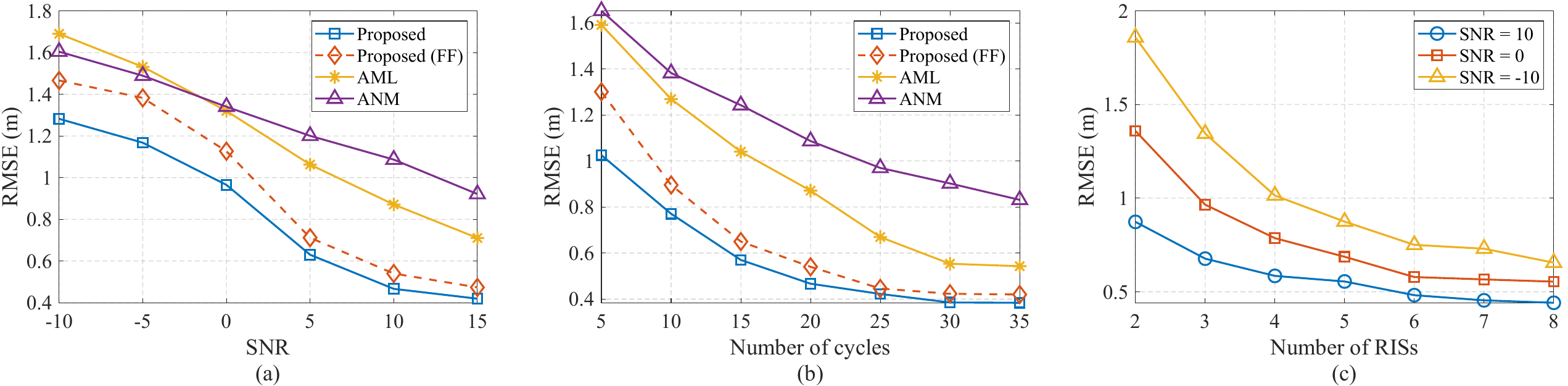}
	\caption{(a) Estimation accuracy of the proposed method for different SNR compared with AML and ANM. (b) Estimation accuracy of the proposed method for different numbers of localization cycles compared with AML and ANM. (c) Estimation accuracy of the proposed method for different numbers of RISs.}
	\label{fig}
\end{figure*}

\section{Simulation Results}
This section presents simulation results validating the performance of our proposed method. We consider a system of $6$ RISs and select $3$ RISs to locate each user. The system operates at a center frequency of $28$GHz. We conduct $T = 500$ independent trials, each with $20$ cycles. For comparison, we include the results of three benchmark approaches: (1) approximate maximum likelihood~(AML)~\cite{b20}, an ML-based NF localization method; (2) atomic norm minimization~(ANM)~\cite{b53}, an off-grid direction-of-arrival estimation method; and (3) the FF implementation of our proposed algorithm~(labeled as FF), where we combine FF model with the proposed method to highlights the benefits of the hybrid model.

Fig.~\ref{fig}~(a) illustrates the average estimation root mean square error~(RMSE) versus the signal-to-noise ratio (SNR). The RMSE is defined as the average localization error, given by $\text{RMSE} = \sqrt{\frac{1}{TK} \sum_{t, k} \Vert \hat{\bm{p}}_{kt} - \bm{p}_k\Vert^2}$,
where $\hat{\bm{p}}_{kt}$ is the estimated location of the $k$-th user at the $t$-th trail. We can see that the proposed algorithm with the hybrid model performs better than with the FF model. This is because the FF implementation cannot provide high localization accuracy for NF users due to model mismatch. Fig.~\ref{fig}~(b) shows the RMSE versus the number of cycles when SNR = 10. As observed, the RMSE initially decreases as the number of cycles increases, but stops improving when the number of cycles is very large. This is attributed to two factors. First, the estimated location is constrained to the sampled grids, while the actual user locations are off-grid. Second, the algorithm might not reach the nearest grid point due to possible local minima.


From Fig.~\ref{fig}~(a) and Fig.~\ref{fig}~(b), it is evident that the proposed algorithm outperforms other algorithms, demonstrating its superiority. The main reasons for this phenomenon are as follows: The AML is an approximated ML method that iteratively updates the range and angle. This approximation has a negative impact on the localization accuracy. The ANM and the FF scheme are based on the plane wave model, which fails to describe the NF channel accurately.

Fig.~\ref{fig}~(c) illustrates the influence of the number of RIS. We vary the number of RISs while maintaining the RIS selection percentage as 50\%. Although increasing the number of RISs could increase interference, our method effectively mitigates this issue and further improves accuracy. Specifically, the proposed phase shift optimization algorithm can suppress inter-R IS interference. Moreover, the diversity gain provided by each additional RIS enhances localization accuracy through improved independent localization and the result fusion.

\section{Conclusion}
\label{sec:con}
This letter presents a multi-user localization method for hybrid NF and FF scenarios utilizing multiple RISs. A two-stage localization algorithm: a local estimation step and a fusion-based global localization refinement, was proposed. We derived the CRB of the estimated location, and an RIS phase shift optimization method based on ADMM was developed. Simulation results demonstrate three key advantages of the proposed method: 1) the proposed method reduces RMSE by over 40\% compared to baseline algorithms, achieving sub-meter localization accuracy; 2) the hybrid model demonstrates superior performance to the pure FF model; 3) the localization accuracy can be improved by deploying more RISs.

\begin{appendices}

\end{appendices}

\bibliographystyle{IEEEtran}
\bibliography{IEEEabrv,ref}

\begin{thebibliography}{10}
\providecommand{\url}[1]{#1}
\csname url@samestyle\endcsname
\providecommand{\newblock}{\relax}
\providecommand{\bibinfo}[2]{#2}
\providecommand{\BIBentrySTDinterwordspacing}{\spaceskip=0pt\relax}
\providecommand{\BIBentryALTinterwordstretchfactor}{4}
\providecommand{\BIBentryALTinterwordspacing}{\spaceskip=\fontdimen2\font plus
\BIBentryALTinterwordstretchfactor\fontdimen3\font minus \fontdimen4\font\relax}
\providecommand{\BIBforeignlanguage}[2]{{%
\expandafter\ifx\csname l@#1\endcsname\relax
\typeout{** WARNING: IEEEtran.bst: No hyphenation pattern has been}%
\typeout{** loaded for the language `#1'. Using the pattern for}%
\typeout{** the default language instead.}%
\else
\language=\csname l@#1\endcsname
\fi
#2}}
\providecommand{\BIBdecl}{\relax}
\BIBdecl

\bibitem{b75}
H.~Q. Ngo, A.~Ashikhmin, H.~Yang, E.~G. Larsson, and T.~L. Marzetta, ``Cell-free massive {MIMO} versus small cells,'' \emph{IEEE Trans. Wirel. Commun.}, vol.~16, no.~3, pp. 1834--1850, Mar. 2017.

\bibitem{b76}
Y.~Zhang, B.~Di, H.~Zhang, J.~Lin, C.~Xu, D.~Zhang, Y.~Li, and L.~Song, ``Beyond cell-free {MIMO}: Energy efficient reconfigurable intelligent surface aided cell-free {MIMO} communications,'' \emph{IEEE Trans. Cogn. Commun. Netw.}, vol.~7, no.~2, pp. 412--426, Jun. 2021.

\bibitem{b72}
R.~Deng, B.~Di, H.~Zhang, Y.~Tan, and L.~Song, ``Reconfigurable holographic surface-enabled multi-user wireless communications: Amplitude-controlled holographic beamforming,'' \emph{IEEE Trans. Wirel. Commun.}, vol.~21, no.~8, pp. 6003--6017, Aug. 2022.

\bibitem{b69}
J.~He, A.~Fakhreddine, H.~Wymeersch, and G.~C. Alexandropoulos, ``Compressed-sensing-based {3D} localization with distributed passive reconfigurable intelligent surfaces,'' in \emph{Proc. IEEE Int. Conf. Acoust. Speech Signal Process. (ICASSP)}, Rhodes Island, Greece, Jun. 2023.

\bibitem{b70}
Y.~Zhang, Y.~Liu, Y.~Liu, L.~Wu, Z.~Zhang, and J.~Dang, ``Multi-{RIS}-assisted millimeter wave single base station localization,'' in \emph{Proc. Inf. Commun. Technol. Conf. (ICTC)}, Nanjing, China, May 2023.

\bibitem{b78}
S.~Zhao, Y.~Liu, L.~Wu, J.~Rodríguez-Piñeiro, X.~Yin, and J.~Hong, ``Recursive {UE} localization for a multi-{RIS}-assisted wireless system in an obstacle-dense environment,'' in \emph{Proc. Eur. Conf. Antennas Propag. (EuCAP)}, Mar. 2023, pp. 1--5.

\bibitem{b80}
B.~Di, H.~Zhang, Z.~Han, R.~Zhang, and L.~Song, ``Reconfigurable holographic surface: A new paradigm for ultra-massive {MIMO},'' \emph{IEEE Trans. Cogn. Commun. Netw.}, early access, Mar. 2025.

\bibitem{b36}
J.~He, A.~Fakhreddine, C.~Vanwynsberghe, H.~Wymeersch, and G.~C. Alexandropoulos, ``3{D} localization with a single partially-connected receiving {RIS}: Positioning error analysis and algorithmic design,'' \emph{IEEE Trans. Veh. Technol.}, vol.~72, no.~10, pp. 13\,190--13\,202, May 2023.

\bibitem{b6}
F.~Zhang, M.-M. Zhao, M.~Lei, and M.~Zhao, ``Joint power allocation and phase-shift design for {RIS}-aided cooperative near-field localization,'' in \emph{Proc. Int. Symp. Wirel. Commun. Syst. (ISWCS), Hangzhou, China}, Oct. 2022.

\bibitem{b20}
C.~Ozturk, M.~F. Keskin, H.~Wymeersch, and S.~Gezici, ``\BIBforeignlanguage{English}{{RIS}-aided near-field localization under phase-dependent amplitude variations},'' \emph{\BIBforeignlanguage{English}{IEEE Trans. Wirel. Commun.}}, vol.~22, no.~8, pp. 5550 -- 5566, Apr. 2023.

\bibitem{b77}
\BIBentryALTinterwordspacing
M.~Cao, H.~Zhang, Y.~C. Eldar, and H.~Zhang, ``Hybrid near-field and far-field localization with holographic {MIMO},'' 2025. [Online]. Available: \url{https://arxiv.org/abs/2501.17868}
\BIBentrySTDinterwordspacing

\bibitem{b34}
X.~Wei and L.~Dai, ``Channel estimation for extremely large-scale massive {MIMO}: Far-field, near-field, or hybrid-field?'' \emph{IEEE Commun. Lett.}, vol.~26, no.~1, pp. 177--181, Jan. 2022.

\bibitem{b79}
G.~C. Alexandropoulos, I.~Vinieratou, and H.~Wymeersch, ``Localization via multiple reconfigurable intelligent surfaces equipped with single receive {RF} chains,'' \emph{IEEE Wirel. Commun. Lett.}, vol.~11, no.~5, pp. 1072--1076, May 2022.

\bibitem{b66}
H.~Xu, R.~Liu, Y.~Xie, J.~Li, P.~Zhu, and D.~Wang, ``Cross-region fusion and fast adaptation for multi-scenario fingerprint-based localization in cell-free massive {MIMO} systems,'' \emph{IEEE Wirel. Commun. Lett.}, vol.~13, no.~10, pp. 2882--2886, Oct. 2024.

\bibitem{b59}
L.~Ran, B.~Sun, S.~Chen, and F.~Xi, ``Transmit beampattern synthesis for {RIS}-aided radar via sidelobe level minimization,'' in \emph{Proc. IEEE Int. Conf. Signal Process. Syst. (ICSPS)}, Online, Nov. 2022.

\bibitem{b14}
H.~Chen, Y.~Bai, Q.~Wang, H.~Chen, L.~Tang, and P.~Han, ``{DOA} estimation assisted by reconfigurable intelligent surfaces,'' \emph{IEEE Sens. J.}, vol.~23, no.~12, pp. 13\,433--13\,442, May. 2023.

\bibitem{b53}
P.~Chen, Z.~Yang, Z.~Chen, and Z.~Guo, ``Reconfigurable intelligent surface aided sparse {DOA} estimation method with non-{ULA},'' \emph{IEEE Signal Process. Lett.}, vol.~28, pp. 2023--2027, Sep. 2021.

\end{thebibliography}

\end{document}